\documentclass{article}

\usepackage[english]{babel}

\usepackage[letterpaper,top=2cm,bottom=2cm,left=3cm,right=3cm,marginparwidth=1.75cm]{geometry}

\usepackage{amsmath}
\usepackage{graphicx}
\usepackage{subfig}
\usepackage{amssymb}
\usepackage{braket}
\usepackage{comment}
\usepackage[colorlinks=true, allcolors=blue]{hyperref}
\usepackage{algorithm}
\usepackage{algpseudocode}
\usepackage{caption} 
\usepackage{subfig}
\usepackage{titling}
\usepackage{lipsum}
\title{Robust Estimation of Surface Curvature Information from Point Cloud Data}
\author{Jared Spang \thanks{{Department of Applied Physics and Applied Mathematics, Columbia University, New York, NY 10027; jared.spang@columbia.edu}}}
\date{\small May 10, 2023}

\begin{document}
\maketitle
\begin{abstract}
This paper surveys and evaluates some popular state of the art methods for algorithmic curvature and normal estimation. In addition to surveying existing methods we also propose a new method for robust curvature estimation and evaluate it against existing methods thus demonstrating its superiority to existing methods in the case of significant data noise. Throughout this paper we are concerned with computation in low dimensional spaces ($N < 10$) and primarily focus on the computation of the Weingarten map and quantities that may be derived from this; however, the algorithms discussed are theoretically applicable in any dimension. One thing that is common to all these methods is their basis in an estimated graph structure.  For any of these methods to work the local geometry of the manifold must be exploited; however, in the case of point cloud data it is often difficult to discover a robust manifold structure underlying the data, even in simple cases, which can greatly influence the results of these algorithms. We hope that in pushing these algorithms to their limits we are able to discover, and perhaps resolve, many major pitfalls that may affect potential users and future researchers hoping to improve these methods.  

\end{abstract}
\section{Introduction}

The robust and accurate estimation of surface curvature information from point cloud data is a crucial task in many fields such as computer graphics, robotics, and manufacturing. For example, in the realm of 3D surface analysis we would ideally work with mesh data; however, oftentimes the only information available is that of point cloud data from which we must then estimate the underlying surface / mesh \cite{DBLP:journals/corr/abs-2001-07884}. A particularly salient example demonstrating the need for estimates associated with point clouds is that of LiDAR technologly. LiDAR is a popular remote sensing technology that generates data in the form of point clouds by utilizing light pulses to estimate the distance to a target surface which is currently being used in a variety of applications from autonomous vehicle navigation to atmospheric science \cite{DBLP:journals/corr/abs-1910-13122, Koch2004C02}.

Although there are numerous examples of instances in which we lack explicit surface information, the precise location of highly curved regions can be difficult to determine from just point cloud data; however, there have been great strides made in this field of curvature estimation in the last few decades. A particularly interesting challenge associated with point cloud data is the inherent uncertainty associated with the data. We might imagine a scenario in which the point cloud data at hand was scanned from an inherently noisy sensor and, as a result, there is great uncertainty in the location of the points themselves. In addition to this common complication there are often numerous parameters to tune in manifold learning models which effectively introduce additional noise to the algorithm as there is no inherent "best" parameter choice. 

As a result, designing an algorithm capable of reliably and robustly estimating surface curvature from nothing but point cloud data is an incredibly difficult task. There have been several papers which have sought to solve this problem; however, few provide strict guarantees on notions such as error or convergence. It is worth noting that there have been several papers that take a geometric measure theory approach to solving similar problems which are typically able to prove convergence; however, many of these results do not hold in the case of highly irregular surfaces and are still highly sensitive to the issues discussed \cite{chazal2008stability, cohen2006,Buet2018}.

In the field of geometry it is well known that there are a slew of different notions of curvature; however, here we will be primarily concerning ourselves with the the computation of the Weingarten map and principle curvatures. Specifically, we will be implementing, building on, and analyzing the results of \cite{Cao_2021}  where we directly estimating the Weingarten map and of \cite{merigotVORONOI2011} which performs the estimation of curvature related quantities through the calculation of Voronoi Covariance Measure. The benefit of these methods are that they claim robustness and are able to perform  estimation in arbitrary dimension. There are of course numerous other methods for computing curvature (see \cite{MAGID2007139}); however, most of these methods are specific to $\mathbb{R}^3$. It is our hope that the methods discussed are diverse enough, both methodologically and temporally, to provide an interesting avenue for comparison. It is the purpose of this paper to evaluate the hardiness of these methods and work to evaluate some potential improvements on them.   

\section{Background}

The underlying assumption we will be working with is one common to the field of manifold learning; although accumulated data may be presented in a high dimensional Euclidean space $\mathbb{R}^N$, it actually lies within a low dimensional latent manifold $\mathcal{M} \subseteq \mathbb{R}^N$. This manifold $\mathcal{M}$ is the surface from which we hope to determine our curvature information. Here we will review the necessary mathematical background for the understanding of the algorithms to be presented. We start by defining the Weingarten map and its relationship to the second fundamental form as well as its connection to the mean and Gaussian curvature. As the algorithms discussed in this paper are theoretically viable for arbitrary dimension we present these concepts in their full generality using notions from differential geometry. Following this we develop the theory surrounding Voronoi cells and both establish the Voronoi Covariance measure and its relevant variants. Before we proceed, however, we must establish some ubiquitous notation. Throughout this paper we let $\mathcal{M} \subseteq \mathbb{R}^N$ be our $m$-dimensional manifold which is embedded in $N$-dimensional space. Hence, for any point $x_i \in \mathcal{M}$ our tangent space $T_{x_i}$ is $m$-dimensional. Here we are following the manifold hypothesis where $\mathcal{M}$ is the "low" dimensional manifold in which the data rests.

\subsection{Curvature Estimation and PCA}
When solving the problem of curvature estimation one often first simplifies the problem to that of estimating normal and tangent spaces. This makes intuitive sense since, in $\mathbb{R}^3$, for any particular point $x_i \in \mathcal{M}$ we may associate a normal vector $n$ to that point. Since curvature is a metric of how much a surface curves in a different directions we can estimate the curvature by observing how the surface normal changes across the surface. For example, if a small perturbation $\varepsilon$ in $x_i$ s.t. $x_i + \varepsilon \in \mathcal{M}$ yields a large change in the surface normal then, intuitively, we would hope the surface to be highly curved at that point. So, one can imagine a valid method of determining curvature is to compute the surface normals and observe how they change.   
One common method of determining the surface normals and tangent space is through that of principal component analysis (PCA). The basic algorithm can be defined as follows:

\begin{algorithm}
\caption{PCA Normal and Tangent Estimation}\label{alg:pca_normal}
\begin{algorithmic}[1]
\Function{PCA\_NT\_Est}{$\varepsilon$, $x_i$}
\State $N(x_i) = \{x : \|x-x_i\| < \varepsilon\}$
\State $C_{x_i} = \sum_{x_j \in N(x_i)} (x_j - \Bar{x}_i) \otimes (x_j - \Bar{x_i})$ \Comment{$\Bar{x_i}$ is the mean of $x_j \in N(x_i)$}
\State $\vec{\lambda},E = \text{Eig}(C_{x_i})$ \Comment{$E$ is the matrix of eigenvectors}
\State $n = \{E_1, E_2, \ldots E_m\}$ \Comment{First $m$ eigenvectors}
\State $t = \{E_m, E_{m+1}, \ldots E_N\}$ \Comment{Last $N-m$ eigenvectors}
\State \Return $n$, $t$
\EndFunction
\end{algorithmic}
\end{algorithm}

Here it is worth noting that the eigenvalues $\vec{\lambda}$ are sorted from smallest to largest so the normal space is eigenvectors corresponding to the first $m$ smallest eigenvalues and the tangent space consists of the eigenvectors corresponding the $N-m$ largest eigenvalues. So with this algorithm we have effectively associated an estimated tangent and normal space to each point $x_i$ which can then be used in the curvature computation.  

\subsection{The Weingarten Map and Curvature}
The Weingarten map, also known as the shape operator or second fundamental form, is a mathematical quantity of focus in this paper. Before we define the Weingarten map we first endow our manifold $\mathcal{M}$ with some additional structure; let our manifold $\mathcal{M}$ be a Riemannian manifold $(\mathcal{M}, d)$ endowed with the standard Levi-Civita connection. We then define $\mathcal{S}$ to be the Weingarten map which serves to associate a linear map with each point $x_i \in \mathcal{M}$ on a surface that details how the surface curves in the ambient space. Specifically, the Weingarten map tells us how a normal vector transforms as we move along a chosen tangent vector. Let $x_i$ be a point on $\mathcal{M}$ and $v \in T_{x_i} \mathcal{M}$ then we may more precisely define $\mathcal{S}$ at point $x_i$ as 
\begin{equation}
    \mathcal{S}_{x_{i}} : T_{x_i} \mathcal{M} \mapsto T_{x_i} \mathcal{M}
\end{equation}
\begin{equation} \label{shape_op_def}
    \mathcal{S}_{x_{i}} (v) = -\nabla_v \zeta(x_i) 
\end{equation}
where $\zeta$ is a normal vector field to $\mathcal{M}$ and $\nabla_v$ is the covariant derivative. It is also useful to write the Weingarten map strictly in terms of the connection. For an arbitrary immersion $\imath: \mathcal{M} \mapsto A^N$ which maps an $n$-dimensional manifold $\mathcal{M}$ into an ambient space $A$ we let $\bar{\nabla}$ be the connection on $A$ and $\nabla$ be the induced Riemannian connection on $M$. Then for any tangent fields $X,Y$ to $\mathcal{M}$ we may decompose the connection in terms of tangential ($\top$) and normal ($\bot$) projections \cite{spivak_2005} 
\begin{equation}
    \begin{gathered}
    \bar{\nabla}_{X}Y = \top \bar{\nabla}_X Y + \bot \bar{\nabla}_XY \\
    \bar{\nabla}_{X}Y = \nabla_{X}Y + \mathrm{I\!I}(X,Y)
    \end{gathered}
\end{equation}
where $\mathrm{I\!I}$ is the second fundamental form which serves to define the Weingarten map as
\begin{equation}
    \mathrm{I\!I}(v,w) = \langle \mathcal{S} (v), w \rangle \zeta = \langle -\nabla_v \zeta, w \rangle \zeta 
\end{equation}
So we see the mathematics supports our initial intuition that the change in the normal vector is essential in defining the Weingarten map and curvature. In fact, the relation between the Weingarten map and curvature can be made clear by consider both mean curvature and Gaussian curvature.

In the case of mean curvature we are interested in the average curvature of the surface at a given point which is given by the average of the principal curvatures at the desired point. Similar to mean curvature, we can compute the Gaussian curvature as the product of the principal curvatures at a given point. Luckily we can easily express both of these curvatures in terms of the Weingarten map  
\begin{equation} \label{curvature_comp}
    \begin{gathered}
        H_{x_i} = \frac{1}{n} \text{tr} (S_{x_i}) \\
        K_{x_i} = \det(S_{x_i})
    \end{gathered}
\end{equation}
where $H$ is the mean curvature and $K$ is the Gaussian curvature. Additionally, the directions of principal curvature can be computed as the eigenvectors of $S_{x_i}$. As a result, computation of the Weingarten map $S_{x_i}$ allows us to easily compute both the mean curvature and the Gaussian curvature. 

A final fact of note is the relationship between the Weingarten map and the Gauss map which, for any manifold of dimension $N-1$, maps $x_i \in \mathcal{M}$ to the unit hypersphere $\mathbb{S}^{N-1} \subseteq \mathbb{R}^N$. Intuitively this map allows us to associate a normal $\zeta(x_i)$ with each point $x_i \in \mathcal{M}$. Since we have shown the Weingarten map to be the covariant derivative of the normal field we immediately see the connection between the Gauss map $g$ and $S$,
\begin{equation}
    \begin{gathered}
            g_{x_i} :  T_{x_i} \mathcal{M} \mapsto \mathbb{S}^{N-1} \\
            S_{x_i}(X) = -Dg_{x_i}(X) = \top(-D g_{x_i}(X)) = -\nabla_{X} \zeta (x_i)
    \end{gathered}
\end{equation}
Where $X \in T_{x_i} \mathcal{M}$ and so we have explicitly recovered our equation \eqref{shape_op_def} for the Weingarten map as the differential of the Gauss map. 

\subsection{Voronoi Cells}
Let $K \subset \mathbb{R}^N$ be a general compact subset of $R^N$ then we define the distance metric $d_K$ for any $x \in \mathbb{R}^K$ as
\begin{equation}
    d_K(x) = \min_{y\in K} \|x-y\|
\end{equation}
which just measures the distance to the closest point in $K$ from $x$. Now let $K \subset \mathbb{R}^N$ be a finite set of points (a point cloud) rather than a general compact set; then we define the Voronoi cell, induced by a particular norm $\|\cdot\|$, of point $x_i$ to be 
\begin{equation}
    \text{Vor}(x_i) = \{y: d_K(y) = \|x_i - y\| \}
\end{equation}
which is the set of points in the ambient space $\mathbb{R}^N$ that are closer to $x_i \in K$ than to any other $x_j \in K$. It is worth noting that $\{\text{Vor}(x_i)\}$ for all $x_i \in K$ forms a partition of the space $R^N$ which is identical to that of the K-NN algorithm as Voronoi cells are fundamentally solving the same problem. It is also useful to define the medial axis $\mathbb{M}(K)$ of $K$. The medial axis is the set of points that have more than one nearest Voronoi point and forms the "skeleton" of the Voronoi cell diagram. The key difference between the medial axis and the general boundary of the Voronoi cell is that the medial axis must be equidistant to two or more of the Voronoi points.

From this definition of medial axis we also define the projection function $p_K : \mathbb{R}^N \setminus \mathbb{M}(K) \mapsto K$ which projects any $x \in \text{Vor}(x_i) \subseteq \mathbb{R}^N$ to the closest point in $K$ (in our case $K$ will be our point cloud). The projection function will be used as in \cite{merigotVORONOI2011} for which the authors note that the function is well behaved almost everywhere which is sufficient for the use case of \eqref{alg:vcm} as we are interested purely in the integral of $p_K$.

\subsection{Voronoi Covariance Measure}
Here we define the Voronoi Covariance Measure (VCM) which is a tensor-valued measure defined for any compact set $K \subseteq R^N$.
\begin{equation} \label{VCM_eq}
    \mathcal{V}_{K,R}(B) = \int_{K^R \cap p_k^{-1}(B \cap K)}(x - p_K(x)) \otimes (x - p_K(x)) dx
\end{equation}
The VCM can be thought of as a generalized covariance matrix since replacing $p_K(x)$ with a point $p$ and adjusting the domain yields the definition of covariance. Instead of a single point $p$ we are now integrating over the Voronoi cells of a "curve" of points. This notion of VCM is good because it encodes normal information; consider the set $\{x_i - p : p_K(x_i) = p, x_i \in K^{R}\}$ which takes the set of points that can be projected to $x_i$ and does so, then this set is the normal cone for the point $x_i$. So for a small $\varepsilon$-ball around $x_i$ the VCM is the sum of the covariance matrices (across an infinitesimally small volume) of the normal cones about that point. Thus, the VCM has a clear relation with the variation in the normal of a surface which helps us intuitively establish a connection to the Weingarten map and the curvature of the surface.   

In addition to the standard VCM we can also convolve the measure by a convolution kernel function $\chi : \mathbb{R}^N \mapsto \mathbb{R}^+$ which allows us to locally smooth the VCM in the case of highly noisy observations. The convolved VCM can be written as,
\begin{equation}
    \mathcal{V}_{K,R}*\chi (p) = \int_{K^R}(x - p_K(x)) \otimes (x - p_K(x)) \chi(x - p_K(x)) dx
\end{equation}
Typically we are interested in performing a convolution over a ball of radius $r$ local to the desired point $x_i$ and as such we set $\chi$ to be the indicator function of the $r$-ball which yields $\mathcal{V}_{K,R}*\chi (p) = \mathcal{V}_{K,R}(B(p,r))$. This provides a convenient way to add robustness to noise via local smoothing.

As we are not working with continuous sets but rather a discrete point cloud what we are primarily interested in is the discrete analog of these functions. Luckily this is quite straightforward and we get the following for the VCM of a point $x_i$ in a point cloud $K$

\begin{equation}
    \mathcal{V}_{K,R}(\{x_i\}) = \int_{K^R \cap p_k^{-1}(x_i)}(y - x_i) \otimes (y - x_i) dy 
\end{equation}
which amounts computing the covariance of $x_i$ across its Voronoi cell (and within the radius $R$). Additionally, we can extend this definition to any set of discrete points by simply summing them together which leads us to the Convolved VCM for point clouds, 
\begin{equation}
    \begin{gathered}
            \mathcal{V}_{K,R}*\chi (\{x_i\}) = \int_{K^R}(y - x_i) \otimes (y - x_i) \chi(y - x_i) dy \\
            \mathcal{V}_{K,R}*\chi (\{x_i\}) = \sum_{x_j \in B(x_i,r) \cap K} \text{cov}(\text{Vor}(x_j) \cap B(x_j,R), x_j)
    \end{gathered}
\end{equation}
This provides us with a syntactically simple way to express the "smoothed" VCM; however, it can still be difficult to compute the covariance across an arbitrary Voronoi cell which is why we resort to Monte-Carlo integration in \eqref{alg:vcm}.   

\subsection{Theoretical Guarantees of VCM}
A particularly useful facet of VCM based algorithms is that VCM and convolved VCM are provably robust. Essentially, if we consider two sets $K$ and $K^\varepsilon$ where $K^\varepsilon$ is some "noised" version of $K$ then if $K$ and $K^\varepsilon$ are close in Hausdorff distance then they are close in convolved VCM. We can make this formal using the results of \cite{merigotVORONOI2011}. Let $d_H(X,Y)$ denote the Hausdorff distance between two sets $X$ and $Y$. If for every $K \subseteq \mathbb{R}^N$ and $R >0$, $\chi : \mathbb{R}^N \mapsto \mathbb{R}$ is bounded and $k$-Lipschitz then then there exists a constant $C$ s.t. for any other $K^\varepsilon \subseteq \mathbb{R}^N$ 

\begin{equation}
    \| \mathcal{V}_{K,R} * \chi - \mathcal{V}_{K^{\varepsilon},R}*\chi \|_{\infty} \leq C d_H(K,K^\varepsilon)^{1/2}
\end{equation}
Where $C$ is a constant that depends on $N,K$ and $R$. So if the Hausdorff distance between $K$ and $K^\varepsilon$ is $\varepsilon$ we get
\begin{equation}
\| \mathcal{V}_{K,R} * \chi - \mathcal{V}_{K^{\varepsilon},R}*\chi \|_{\infty} \leq C \varepsilon^{1/2}
\end{equation}
Interpreting this result we see that for a fixed compact set (or point cloud) $K$, ambient dimension $N$, and $R$ we have that the maximum difference in the convolved VCM between $K$ and $K^\varepsilon$ is bounded by a constant times the square root of the Hausdorff distance. This essentially let us know that as the Hausdorff distance between two sets shrinks (e.g. because there is less noise) so does the difference in their convolved VCM.   
Although this result proves that the convolved VCM converges in Hausdorff distance, a common complaint lodged against this method is that we are not often faced with "Hausdorff noise" in practice. As a result, some work has been done to generalize this method in a way that provides more robust convergence properties in the case of more realistic noise; however, we will not involve ourselves with these variants in this paper \cite{Cuel_2015,geo_inference_for_measures}.

\section{Numerical Estimators} 

\subsection{Monte-Carlo VCM Estimator}
The Monte-Carlo Voronoi Covariance Measure Estimator (MCVCM) is a randomized algorithm for estimating the Voronoi Covariance Measure. This algorithm has been shown to, with high probability, converge to an $\varepsilon$-approximation of the true VCM when run for $N = \mathcal{O}(|K| \ln(1/\varepsilon) / \varepsilon^2)$ iterations. It's worth highlighting here that the VCM is a tensor field which, in $\mathbb{R}^N$, associates an $N \times N$ matrix with any point $x \in R^N$. So if we wish to compute the VCM for a point cloud $K$ with $P$ points at every point $p \in K$ we will have a tensorial array of size $P\times N\times N$. We also note that throughout this paper we perform various calculations using the MCVCM algorithm \eqref{alg:vcm} and in every experiment we choose $N$ large enough to establish an estimation within a 5\% $\varepsilon$-perturbation of the true solution with high probability.     

\begin{algorithm}
\caption{MCVCM}\label{alg:vcm}
\begin{algorithmic}[1]
\Function{VCM}{$K$, $R$, $N$}
\State $V(x) \gets \underline{0}$  $\forall x \in K$ \Comment{Initialize VCM to zero matrix}
\State $M \gets 0$
\For{$1$ to $N$}
    \State $x \gets$ random point in $K$
    \State $s \gets$ uniformly sampled random point in $B(x,R)$
    \State $k \gets \#(B(x,R) \cap C)$ \Comment{Number of points within radius R of x}
    \State $p' \gets$ NearestNeighbor(x,C)
    \State $V(p') \gets V(p') + \frac{1}{k}(x-p')\otimes(x-p')$
    \State $M \gets M + \frac{1}{k}$ 
\EndFor
\State $V \gets V / M$
\State \Return $V$
\EndFunction
\end{algorithmic}
\end{algorithm}

\subsection{WME Estimator}
Once we are able to accurately estimate the normals and tangent bundle of our point cloud via a method such as \eqref{alg:pca_normal} we can estimate the Weingarten map via the WME estimator \eqref{alg:WME}. The WME estimator can be derived as follows, first consider our manifold $\mathcal{M} \subseteq \mathbb{R}^N$. Let $p \in  \mathcal{M}$ and $q \in \mathcal{M}$ is a point s.t. $d_G(p,q) <  \varepsilon$ where $d_G$ is the geodesic distance on $\mathcal{M}$. As shown in \cite{Cao_2021} we have that
\begin{equation}
    \top (\zeta_p - \zeta_q) = -\mathcal{S}_p (\top (p-q) ) + \mathcal{O}(\|p - q\|^2)
\end{equation}
where $\top$ is again the projection to the tangent space. This result is quite interesting as it allows us to directly model the local change in the Weingarten map as a change in normals. This relation can be derived by considering the exponential map $\bf{r}: \mathbb{R}^N \mapsto \mathcal{M}$ s.t. $\bf{r(0)} = p$ (i.e. the geodesic spawning from $p$). Additionally, let $\bf{u} \in \mathbb{R}^N$ then the Taylor expansion of $\zeta(\bf{r}(u)) - \zeta(\bf{r}(0)) = \zeta(\bf{r}(u)) - \zeta(p)$ results in the following,
\begin{equation} \label{WME_T_SERIES}
    \top (\zeta(\bf{r}(u)) - \zeta(p)) = \top (\sum_{i=1}^N u^i \frac{\partial \zeta}{\partial u^i}) + \mathcal{O}(\|p - q\|)^2 = -\mathcal{S}_p (\top (p-q) ) + \mathcal{O}(\|p - q\|^2)
\end{equation}
We would initially expect the higher order term to involve some square of the exponential map; however, by evaluating the exponential map Taylor expansion it can be shown that this higher order term is on the same order as the euclidean distance \cite{Cao_2021,monera2014taylor}. Consider the tangent basis matrix $E_{x_i} = [e_{x_i}^1,\ldots,e_{x_i}^m]$ then from \eqref{WME_T_SERIES} we have 
\begin{equation}
    (\zeta_{x_j} - \zeta_{x_i})^{\intercal} E_{x_i} = -(x_j - x_i)^{\intercal} E_{x_i} S_{x_i}
\end{equation}
Then we hope to find a matrix $S_{x_i}$ which minimizes the residual of this equation,
\begin{equation} \label{WME_MSE_FORM}
    S_{x_i} \approx \min_{S} \sum_{j=1}^N \chi(|x_j - x_i| < \varepsilon)  \|(\zeta_{x_j} + \zeta_{x_i})^{\intercal}E_{x_i} - (x_j - x_i)^{\intercal}E_{x_i}S\|^2  
\end{equation}
where $\chi(|x_j - x_i| < \varepsilon)$ is the indicator function for if $x_j$ is in the $\varepsilon$ ball of $x_i$. This indicator is necessary as our approximation only holds for points near $x_i$ on the manifold; however, we could replace it with some other measure of closeness such as the K nearest neighbors of $x_i$. Thus, we have derived an approximation of the Weingarten map which is solvable as a fairly simple least-square problem for matrices.     
\begin{algorithm}
\caption{WME}\label{alg:WME}
\begin{algorithmic}[1]
\Function{WME}{$\varepsilon$, $K$, $i$}
\State $\zeta,  E \gets \text{PCA\_NT\_EST}(\varepsilon,K)$ \Comment{Get normal and tangent basis}
\State $\Delta \gets E^\intercal [x_1 - x_i,\ldots,x_n - x_i]$
\State $\Xi \gets E^\intercal [\zeta_{x_1} - \zeta_{x_i}, \ldots, \zeta_{x_n} - \zeta_{x_i}]$
\State $W \gets \text{diag}\{\chi(x_1 - x_i),\ldots,\chi(x_n-x_i)\}$ \Comment{KNN mask matrix}
\State $S = -\Xi W \Delta^\intercal (\Delta W \Delta^\intercal)^{-1}$ \Comment{Closed form to LLS}
\State \Return $S$
\EndFunction
\end{algorithmic}
\end{algorithm}

This algorithm has been theoretically shown to be consistent with bias $\mathcal{O}(h\varepsilon^4)$ and variance $\mathcal{O}(\frac{1}{n\varepsilon^m})$. Here we define $\varepsilon$ as the "bandwidth" of the model, which essentially serves as the size of the neighborhood surrounding $x_i$ we take points from (this is $\varepsilon$ in \eqref{WME_MSE_FORM}). Additionally, we have $m$ as the dimension of the manifold and $n$ as the number of points. Thus it has been shown that 
$$MSE = \mathcal{O}(\varepsilon^4 + \frac{1}{n\varepsilon^m})$$
By considering the formulas for curvature in \eqref{curvature_comp} we can then easily extend computation of $S_{x_i}$ via \eqref{alg:WME} to computation of mean and Gaussian curvature.

\subsection{VWME}
Based on the robustness and superior orientability (see Discussion) associated with VCM normal estimation compared to that of PCA-based normal estimation we propose to modify the WME algorithm by augmenting it with the VCM. Specifically, we replace the PCA-based normal and tangent estimation with the normal and tangent estimation found by computing the eigenbasis of the VCM (or locally convolved VCM in this case). We title this variation on the WME algorithm the Voronoi Weingarten Map Estimator (VWME) and refer to it as such throughout the remainder of the paper. The exact details are shown in \eqref{alg:VWME} where we see VWME amounts to this simple mix of VCM and WME algorithm.       

\begin{algorithm}
\caption{VWME}\label{alg:VWME}
\begin{algorithmic}[1]
\Function{VWME}{$\varepsilon$,$r$,$VCM$,$K$,$i$}
\State $N(x_i) \gets \{x : \|x-x_i\| < \varepsilon\}$ \Comment{$\varepsilon$-ball or KNN}
\State $CVCM \gets \sum_{x_j \in N(x_i)} VCM(x_i)$ \Comment{Compute Convolved VCM}
\State $\vec{\lambda},E \gets \text{Eig}(CVCM)$ \Comment{$E$ is the matrix of eigenvectors}
\State $\zeta \gets \{E_1, E_2, \ldots E_m\}$ \Comment{First $m$ eigenvectors}
\State $E \gets \{E_m, E_{m+1}, \ldots E_N\}$ \Comment{Last $N-m$ eigenvectors}
\State $\Delta \gets E^\intercal [x_1 - x_i,\ldots,x_n - x_i]$
\State $\Xi \gets E^\intercal [\zeta_{x_1} - \zeta_{x_i}, \ldots, \zeta_{x_n} - \zeta_{x_i}]$
\State $W \gets \text{diag}\{\chi(x_1 - x_i),\ldots,\chi(x_n-x_i)\}$ \Comment{KNN mask matrix}
\State $S = -\Xi W \Delta^\intercal (\Delta W \Delta^\intercal)^{-1}$ \Comment{Closed form to LLS}
\State \Return $S$
\EndFunction
\end{algorithmic}
\end{algorithm}

\section{Error Analysis}

\begin{figure}

  \centering
    \subfloat[\centering Torus Sectional]{{\includegraphics[width=7cm]{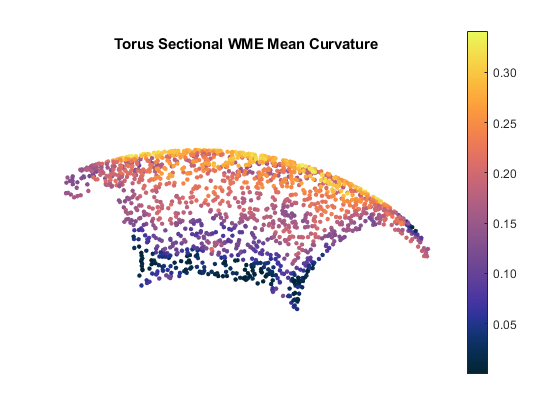} }}%
    \qquad
    \subfloat[\centering Torus]{{\includegraphics[width=7cm]{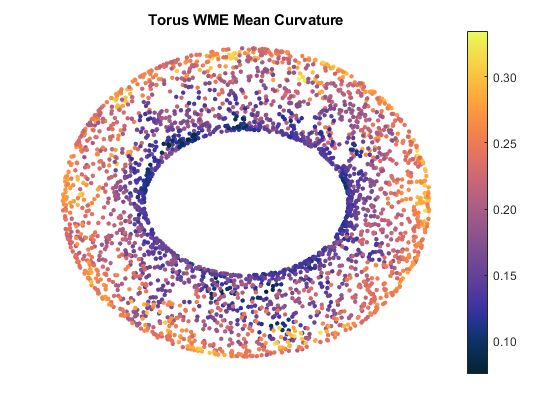} }}%
    \caption{Applying the WME algorithm \eqref{alg:WME} to estimate the mean curvature of the torus (right) and torus sectional (left). This plot demonstrates the baseline capability of the WME algorithm to estimate curvature on some simple 3D shapes with no noise and fairly dense point clouds (2000 points in each example).}%
\label{fig:toroids_mean_curvature}
\end{figure}

\begin{figure}

  \centering
    \subfloat[\centering Torus Sectional]{{\includegraphics[width=7cm]{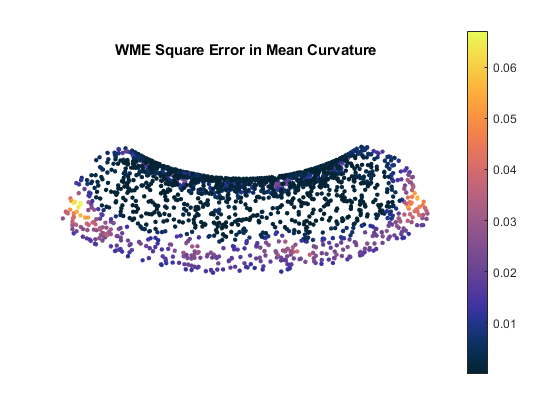} }}%
    \qquad
    \subfloat[\centering Torus]{{\includegraphics[width=7cm]{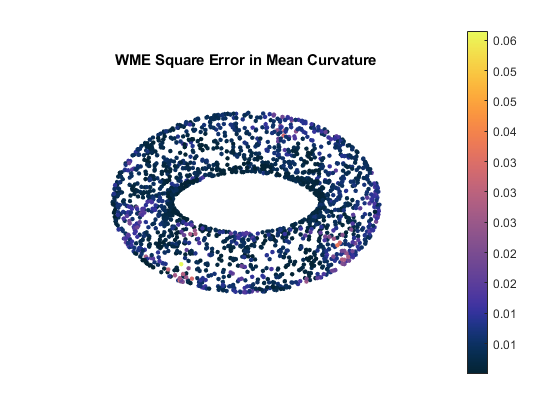} }}%
    \caption{Mean squared error in the WME algorithm \eqref{alg:WME} mean curvature computation of figure \ref{fig:toroids_mean_curvature}. Again we have 2000 points and here we have rotated the view of the to get a better view of high error areas. We see that in the case of the torus the errors are fairly randomly distributed across the surfaces; however, in the torus sectional we get localized errors around the edges of the surface as we may expect. }%
\label{fig:toroids_mean_curvature_error}
\end{figure}

\subsection{Normal Estimators}
As many algorithms for curvature estimation utilize the change in the normals for that estimation it is nature to evaluate the estimation of the surface normals prior to that of the entire algorithm. Luckily for many simples surfaces in $\mathbb{R}^3$ we can construct point clouds with known normal information which allows us to compare our estimates of the normal basis to that of our known surfaces. The most common method for estimating normals is that of the PCA based method detailed in \eqref{alg:pca_normal} which we also evaluate. In addition to the PCA estimator one of the primary use cases propounded for the Voronoi Covariance Measure is as an alternative method for estimating surface normals. As a result, we thoroughly compare the robustness of these two algorithms and how they compare to the true normals in known surfaces.

\begin{figure}

  \centering
    \subfloat[\centering Cosine Similarity]{{\includegraphics[width=7cm]{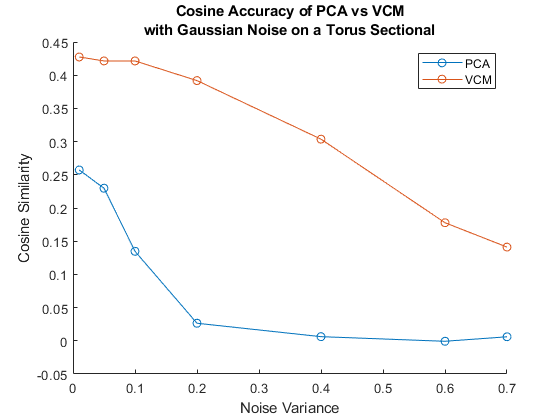} }}%
    \qquad
    \subfloat[\centering Absolute Cosine Similarity]{{\includegraphics[width=7cm]{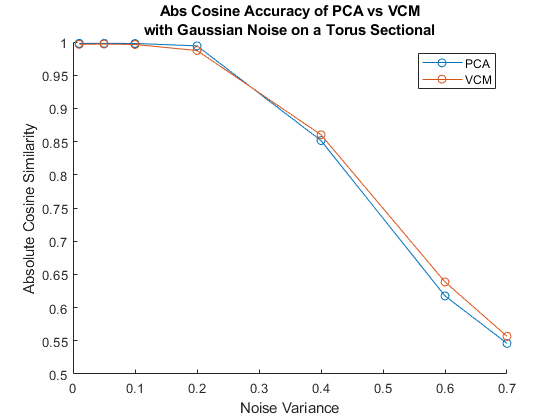} }}%
    \caption{Cosine and Absolute Cosine Similarity between ground truth and the PCA \eqref{alg:pca_normal} and VCM \eqref{alg:vcm} normal estimators. Here we evaluate the models on the torus sectional point cloud with 1000 points with increasing additive Gaussian noise generated with variance $\sigma^2$ (x-axis). The parameters for the convolved VCM estimator are $r=0.2$ and $R=0.5$. }%
\label{fig:pca_vcm_gaussian_cosine_torus_sectional}
\end{figure}

In the case of computing normal vectors we have little interest in the magnitude of the vectors but rather of its angle as we are concerned with the proper decomposition of the local space around a point $x$ into normal and tangent spaces. In order to accommodate this requirement we compute a proxy for the error (or lack thereof) as the cosine similarity in the computed normals to that of the true normals. In the case of increasing Gaussian noise applied to a sectional torus point cloud, we plot both the mean cosine similarity and the mean absolute cosine similarity in figure \ref{fig:pca_vcm_gaussian_cosine_torus_sectional}. Specifically, to determine the cosine similarity ($S_C(A,B)$) between two vectors $A$ and $B$ we compute $S_C(A,B) = \frac{\langle A, B \rangle}{\|A\|\|B\|}$ and the absolute cosine similarity as $|S_C(A,B)|$.

\begin{figure}
    \centering
    \includegraphics[width=0.7\textwidth ]{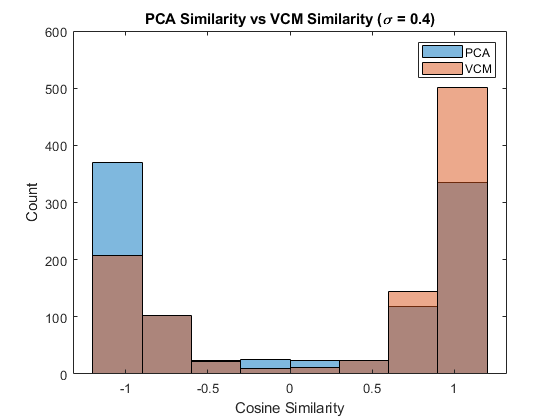}
    \caption{Here we plot a histogram of the Cosine Similarity between the PCA and VCM normal estimators under moderate additive Gaussian noise ($\sigma = 0.4$). Here we use 50-NN for both PCA and VCM on the torus sectional point cloud with 1000 points where the convolved VCM uses the parameters $r=0.2$ and $R=0.5$.}
    \label{fig:pca_vcm_similarity_histogram_sigma04}
\end{figure}

The reason for the inclusion of both the absolute and regular cosine similarity is that when viewing the distribution of errors for both these methods, e.g. in figure \ref{fig:pca_vcm_similarity_histogram_sigma04}, we see that a large number of errors crop up from simply orienting the normal incorrectly. Merely orienting the normal in the wrong direction is not usually a huge issue in practice as the vector will still be orthogonal to the tangent space. In addition, there exist efficient algorithms for establishing a consistent normal orientation across a surface using minimal spanning trees \cite{hoppe1992surface}.

\begin{figure}

  \centering
    \subfloat[\centering Cosine Similarity]{{\includegraphics[width=7cm]{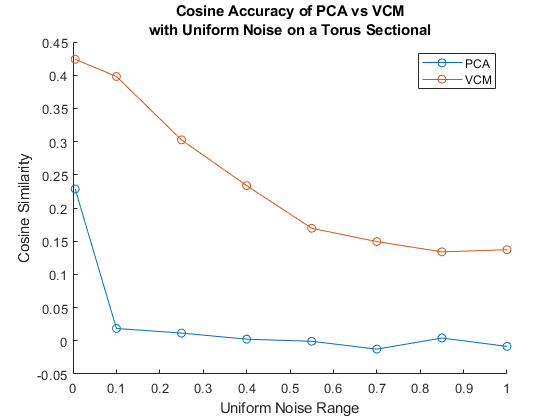} }}%
    \qquad
    \subfloat[\centering Absolute Cosine Similarity]{{\includegraphics[width=7cm]{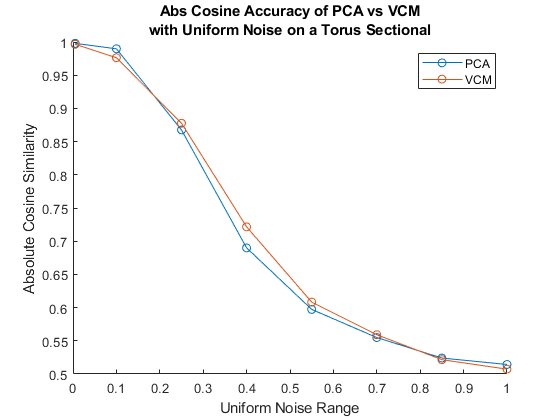} }}%
    \caption{Cosine and Absolute Cosine Similarity between ground truth and the PCA \eqref{alg:pca_normal} and VCM \eqref{alg:vcm} normal estimators. Here we evaluate the models on the torus sectional point cloud with 1000 points with increasing additive uniform noise generated in a range $[-\alpha, \alpha]$ for a parameter $\alpha$ (x-axis). The parameters for the convolved VCM estimator are $r=0.2$ and $R=0.5$. }%
\label{fig:pca_vcm_uniform_cosine_torus_sectional}
\end{figure}

\subsection{WME Estimator}
Although \cite{Cao_2021} does perform a series of error analyses through a variation in the number of sample points on the surface not much is done in the way of varying the construction of the approximation to the manifold itself. As a result, we benchmark the performance of this algorithm under various changes in the type of neighborhood embedding used (K-NN, $\varepsilon$-ball, and Gaussian kernel) as well as the parameters for each of these embeddings. In addition, the performance of the algorithm under various different types of noise is examined for a variety of surfaces.   

\begin{figure}

  \centering
    \subfloat[\centering Noiseless]{{\includegraphics[width=7cm]{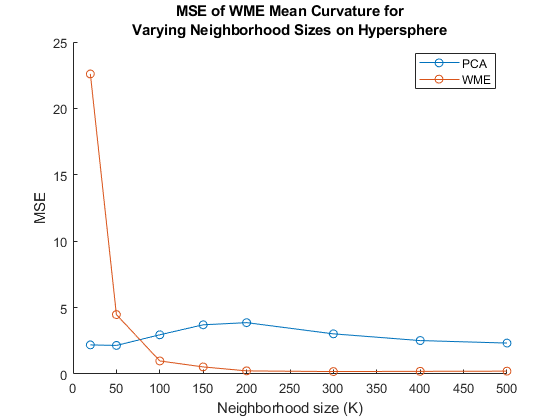} }}%
    \qquad
    \subfloat[\centering Heavy Gaussian Noise ($\sigma = 0.7$)]{{\includegraphics[width=7cm]{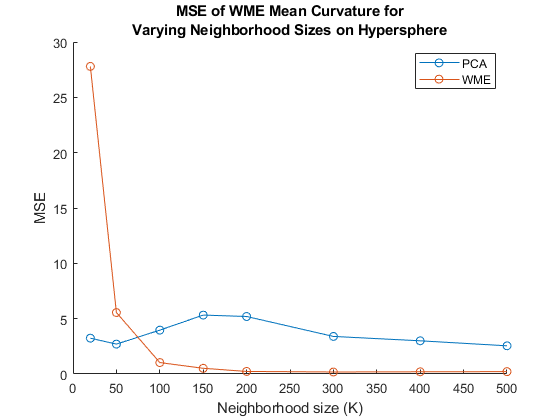} }}%
    \caption{Here we plot the MSE of the computation of the Mean Curvature varying both the PCA neighborhood size and the WME-$\chi$ neighborhood size both approximated via kNN. Here we plot both the noiseless case and the case with $\sigma = 0.7$ additive Gaussian noise. These are both evaluated across the $\mathbb{S}^4$ unit hypersphere sampled with 3000 points. }%
\label{fig:hypersphere}
\end{figure}

One thing that is particularly nice about the WME estimator is that, unlike many estimators used in computer graphics, it is directly applicable to surfaces of arbitrary dimension. As a result, although we cannot visualize it, we can evaluate how the algorithm performs on high dimensional shapes such as the hypersphere. A particular example evaluating the performance of the algorithm on the hypersphere can be seen in figure \ref{fig:hypersphere}.

\begin{figure}
    \centering
    \includegraphics[width=0.85\textwidth ]{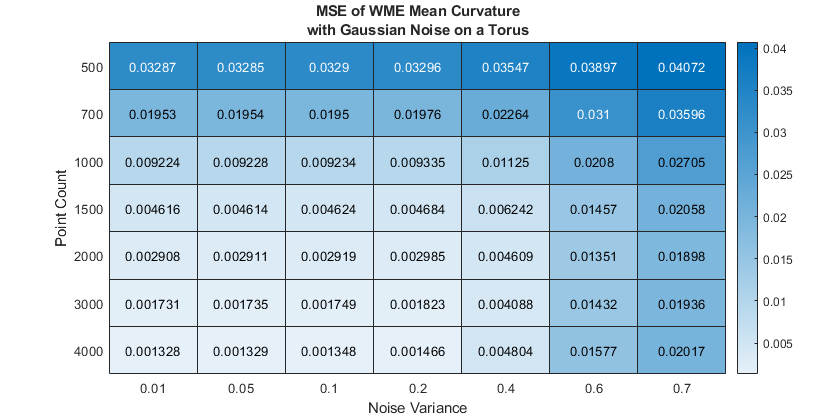}
    \caption{Mean Squared Error (MSE) of Mean Curvature approximation via the WME algorithm \eqref{alg:WME} on the torus when varying both point cloud size and variance of the additive Gaussian noise. For consistency, in every MSE Mean Curvature approximation the PCA \eqref{alg:pca_normal} kNN was computed with 50-NN and the WME-$\chi$ was computed with 30-NN.}
    \label{fig:mse_wme_mcurv_gaussian_torus}
\end{figure}

\begin{figure}
    \centering
    \includegraphics[width=0.85\textwidth ]{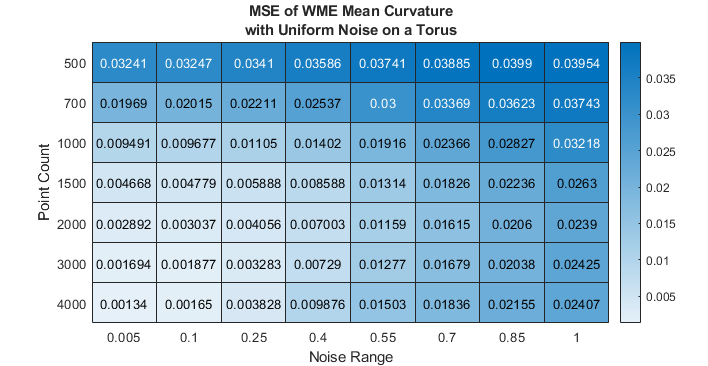}
    \caption{Mean Squared Error (MSE) of Mean Curvature approximation via the WME algorithm \eqref{alg:WME} on the torus when varying both point cloud size and range of additive uniform noise. For consistency, in every MSE Mean Curvature approximation the PCA \eqref{alg:pca_normal} kNN was computed with 50-NN and the WME-$\chi$ was computed with 30-NN.}
    \label{fig:mse_wme_mcurv_uniform_torus}
\end{figure}

\begin{figure}
    \centering
    \subfloat[\centering Noiseless]{{\includegraphics[width=7cm]{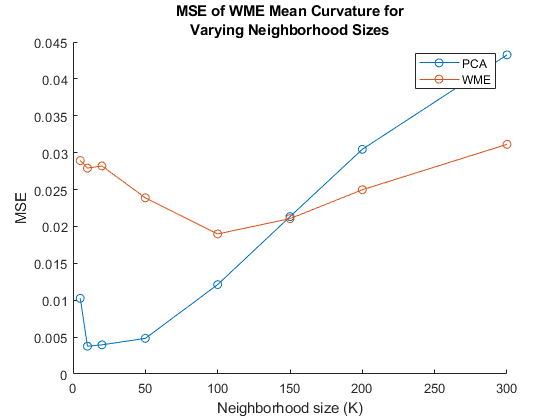} }}%
    \qquad
    \subfloat[\centering Heavy Gaussian Noise ($\sigma = 0.7$)]{{\includegraphics[width=7cm]{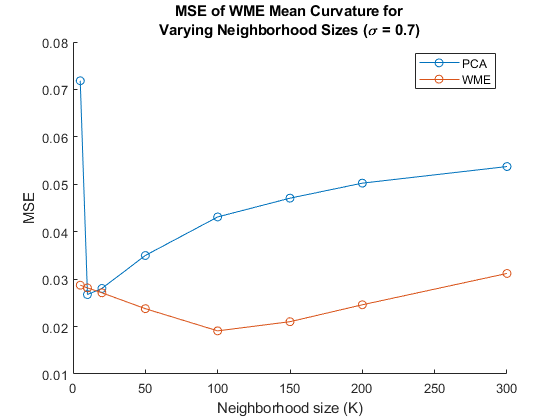} }}%
    \caption{This plot explores the effect on Mean Curvature MSE of modifying the neighborhood size (chosen by KNN) for a torus point cloud with 2000 points in both the of significant ($\sigma = 0.7$) Gaussian noise and the case of no noise. The PCA line depicts the MSE as we vary the kNN size of the PCA algorithm \eqref{alg:pca_normal} whereas the WME line depicts the MSE as we vary the kNN size of the WME-$\chi$ parameter. In both cases when varying the neighborhood size the $k$ parameter the other $k$ value was kept constant at the proposed "optimal" value which is $k = \mathcal{O}(\log n)$ for a set of $n$ points \cite{Cao_2021}. }%
    \label{fig:KNN_neighborhood_varying_MSE_curvature}
\end{figure}

\begin{figure}
    \centering
    \subfloat[\centering Noiseless]{{\includegraphics[width=7cm]{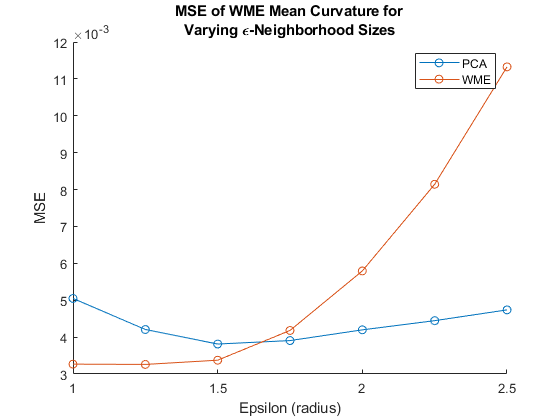} }}%
    \qquad
    \subfloat[\centering  Heavy Gaussian Noise ($\sigma = 0.5$)]{{\includegraphics[width=7cm]{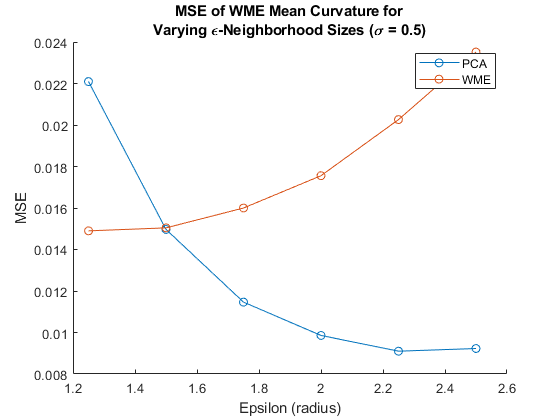} }}%
    \caption{Here we plot just the noiseless case explores the effect on Mean Curvature MSE of modifying the neighborhood size (chosen by KNN) for a torus point cloud with 2000 points in both the of significant ($\sigma = 0.7$) Gaussian noise and the case of no noise. The PCA line depicts the MSE as we vary the kNN size of the PCA algorithm \eqref{alg:pca_normal} whereas the WME line depicts the MSE as we vary the kNN size of the WME-$\chi$ parameter. In both cases when varying the neighborhood size the $k$ parameter the other $k$ value was kept constant at the proposed "optimal" value which is $k = \mathcal{O}(\log n)$ for a set of $n$ points \cite{Cao_2021}. }%
    \label{fig:eps_neighborhood_varying_MSE_curvature}
\end{figure}

\subsection{VWME Estimator}
\begin{figure}
    \centering
    \subfloat[\centering Noiseless]{{\includegraphics[width=7cm]{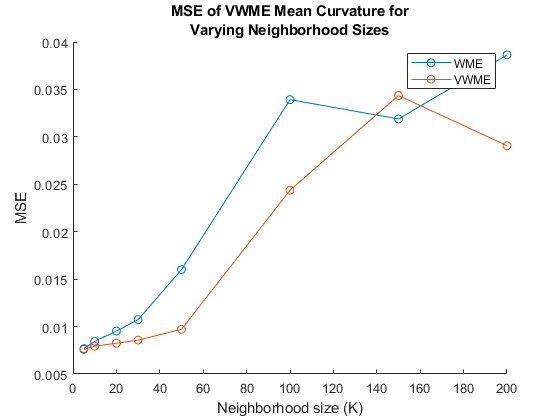} }}%
    \qquad
    \subfloat[\centering  Heavy Gaussian Noise ($\sigma = 0.5$)]{{\includegraphics[width=7cm]{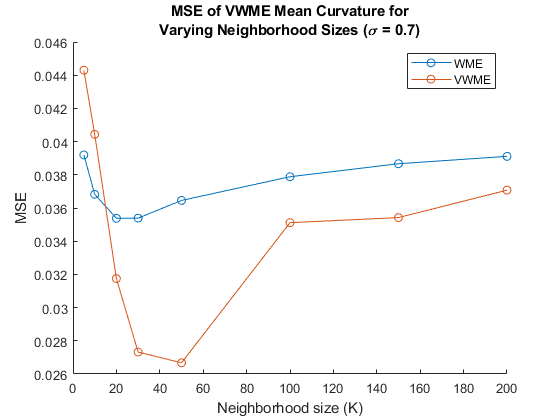} }}%
    \caption{ Here we plot the mean squared error in the calculation of mean curvature using the VWME algorithm \eqref{alg:VWME}. Specifically, we compare the performance of the VWME algorithm to the standard WME algorithm \eqref{alg:WME} as the size of the KNN neighborhood size changes. This comparison is made across a torus point cloud with 1000 points with a VCM \eqref{alg:vcm} R value of 0.5 and a constant VWME-$\chi$ KNN mask with 30-NN. Additionally, this progression was plotted both both the noiseless case and the case of heavy additive Gaussian noise ($\sigma = 0.7$). }%
    \label{fig:mse_of_vwme_mean_curvature_varying_k_noise}
\end{figure}

\begin{figure}
    \centering
    \includegraphics[width=0.85\textwidth ]{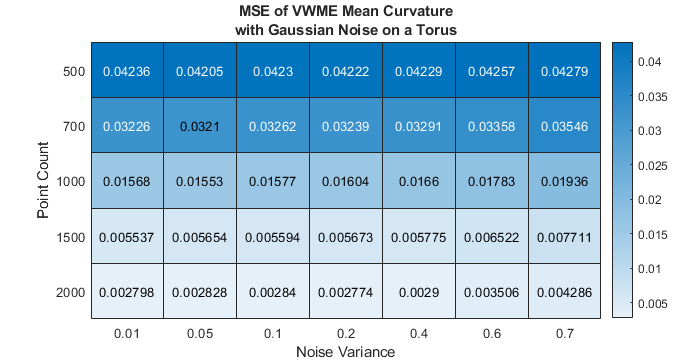}
    \caption{Mean Squared Error (MSE) of Mean Curvature approximation via the VWME algorithm \eqref{alg:VWME} on the torus when varying both point cloud size and range of additive Gaussian noise. For consistency, in every MSE Mean Curvature approximation the VCM \eqref{alg:vcm} kNN was computed with 50-NN and the VWME-$\chi$ was computed with 30-NN.}
    \label{fig:mse_vwme_mcurv_gaussian_torus}
\end{figure}

\begin{figure}
    \centering
    \includegraphics[width=0.85\textwidth ]{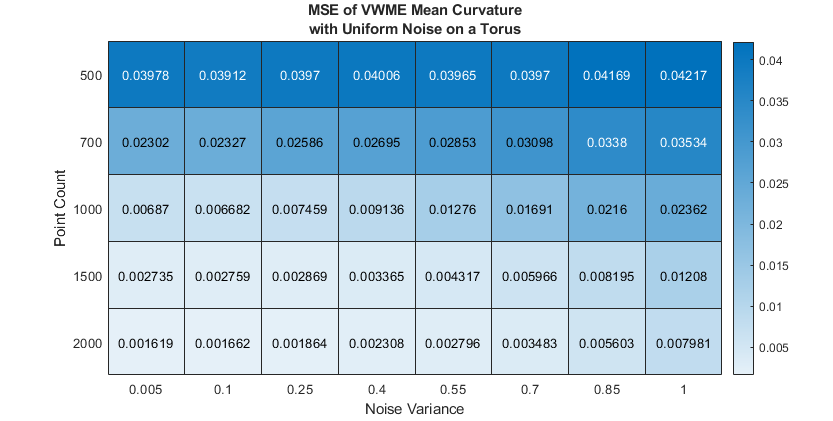}
    \caption{Mean Squared Error (MSE) of Mean Curvature approximation via the VWME algorithm \eqref{alg:VWME} on the torus when varying both point cloud size and range of additive uniform noise. For consistency, in every MSE Mean Curvature approximation the VCM \eqref{alg:vcm} kNN was computed with 50-NN and the VWME-$\chi$ was computed with 30-NN.}
    \label{fig:mse_vwme_mcurv_uniform_torus}
\end{figure}

In addition to performing an error analysis on the WME estimator itself, we also perform an error analysis on our proposed VWME estimator and compare the results to that of the base WME algorithm. Specifically, we evaluate the mean square error in curvature reconstruction across a variety of point cloud sizes with the introduction of increasing additive noise. We consider the case of both Gaussian additive noise (fig \ref{fig:mse_vwme_mcurv_gaussian_torus}) and uniform additive noise (fig \ref{fig:mse_vwme_mcurv_uniform_torus}) just as we have done for the standard WME case. This analysis allows us to compare how robust our proposed VWME algorithm is to noise when compared to the standard WME algorithm.
Not only have we evaluated the VWME algorithm across a variety of noise levels and point cloud sizes but we also directly compare it to the WME algorithm in figure \ref{fig:mse_of_vwme_mean_curvature_varying_k_noise}. Here we consider two cases; a noiseless case and a noised case; in both cases we directly evaluate the VWME algorithm against the WME algorithm when varying the $kNN$ neighborhood size for the normal estimator used. This allows us to directly see how the VWME algorithm holds up against the WME algorithm on a simple test surface.  

\section{Discussion}
As normal / tangent space estimators are the base for our curvature algorithms we began by evaluating these algorithms as in figure \ref{fig:pca_vcm_gaussian_cosine_torus_sectional}. From this graph we see that the introduction of noise in the data certainly has a significant effect on the accuracy (cosine similarity) of the normal estimator. In particular, here we compare the PCA normal estimator to that of the VCM estimator across both cosine similarity and absolute cosine similarity. When viewing these graphs we see that the difference between the two algorithms is much more stark when just comparing cosine similarity. The reason for this can be seen more clearly in figure \ref{fig:pca_vcm_similarity_histogram_sigma04} where we plot the distribution of the cosine similarity for both methods. From this we see that the PCA based estimator performs comparably to the VCM algorithm; however, it has a much greater tendency to flip the direction of the normal thus causing it to yield low overall cosine similarity but decent absolute cosine similarity to the true solution.

We also tested running these algorithms on graphs constructed both by KNN and via $\epsilon$-balls; however, when evaluating noised data on the $\epsilon$-ball graphs a common issue that was encountered was the existence of "singleton" points which were not connected to the rest of the graph. If there is a single singleton point then the algorithms can not be run on that point because they require local information but when the point is disconnected from the main graph there is no local information. One possible way to resolve this would be to generate a more dense sampling of points; however, this is both not always realistic when using real world data and becomes computationally intractable rather quickly. As a result, graphs constructed via KNN over $\epsilon$-balls tend to be superior when one is concerned with the practical application of these algorithms; however, KNN based graphs do not have the same theoretical guarantees as $\epsilon$-balls or kernel graphs (which are even more computationally intractable). One example of this can be seen in figure \ref{fig:eps_neighborhood_varying_MSE_curvature} where smaller $\varepsilon$ values had to be cut from the noised tests. This example is particularly interesting because shows how the addition of noise can directly affect our neighborhood parameter choice. In graph \ref{fig:eps_neighborhood_varying_MSE_curvature}(a) we see a local error minima for the PCA $\varepsilon$ choice around 1.5; whereas, in \ref{fig:eps_neighborhood_varying_MSE_curvature}(b) the error doesn't flatten out until around $\varepsilon = 2.2$ and remains low afterwards. It is possible that further increasing $\varepsilon$ could lead to another rise in error; however, in this particular example the minor radius of the torus is reached at $\varepsilon = 3$ at which point we obtain meaningless results.  

In addition to the tests run the WME algorithm we also ran several error evaluations on our proposed VWME algorithm \eqref{alg:VWME}. As we are mostly concerned with robustness we largely compared the two algorithms in the case of noisy point clouds. After viewing these results we see that, ostensibly, the VWME algorithm outperforms the WME in terms of robustness for sufficiently sampled surfaces ($n \geq 1000$ in the case of figure \ref{fig:mse_vwme_mcurv_gaussian_torus} and figure \ref{fig:mse_vwme_mcurv_uniform_torus}). In fact, as we increase the point count we may compare our VWME results to that of figure \ref{fig:mse_wme_mcurv_gaussian_torus} and figure \ref{fig:mse_wme_mcurv_uniform_torus} and see that the VWME significantly outperforms the WME algorithm in terms of MSE on this surface. Of course, all this demonstrates is that the VWME algorithm has the capability of outperforming the WME algorithm in the case of significant noise. We should keep in mind that the VWME algorithm has an additional parameter choice of $R$ in its normal estimation that the WME algorithm does not have which has the potential to significantly effect the final result. However, in figure \ref{fig:mse_of_vwme_mean_curvature_varying_k_noise} we see that for a fixed $R$ there is a tendency to outperform the WME algorithm when varying the neighborhood size for the normal estimation.        

\section{Conclusion}

Throughout this paper we have developed the theory for several modern normal and curvature estimators that are applicable to point clouds of arbitrary dimension. Following the introduction of relevant theory we presented these algorithms and evaluated them. In particular, we have subjected both PCA-based \eqref{alg:pca_normal} and Voronoi-based \eqref{alg:vcm} curvature estimators to trials of robustness through varying point cloud sizes, parameter sweeping, and a variety of additive noise.  From this we unveiled the behavior of these algorithms and discovered several interesting facets of these algorithms including both the propensity of the PCA estimator to incorrectly orient normals and the finicky nature of $\varepsilon$-ball neighborhood approximations. 
In addition to the evaluation of existing algorithms we also proposed a new variant of the WME algorithm we call VWME \eqref{alg:VWME} which replaces the PCA-based normal estimation of the WME algorithm with a more robust Voronoi-based normal estimation. The intuition behind this new algorithm is that, since the WME algorithm is proven to have low error using true normals, a more robust estimation of the normal vector should yield a more robust overall estimator. We have evaluated this VWME algorithm across a variety of point cloud densities, parameter choices, and noise values and have shown that it is, in general, more robust to noise than the standard PCA-based method for curvature estimation. 

Although we show this algorithm to be more robust some future work still needs to be done. In particular, the algorithm should be tested across a larger and more complex variety of surfaces sampled with a greater number of points. We currently use a Monte-Carlo variation of the VCM algorithm \eqref{alg:vcm}; however, there exists a much more efficient method for computing the VCM involving 3D Delaunay triangulation which could be implemented and used to run these tests \cite{merigotVORONOI2011}. This would be quite beneficial as the VWME algorithm takes significantly longer to run than the WME algorithm and does not scale as well due to its reliance on the VCM. We note that although an efficient variation of VCM exists we stuck with the Monte-Carlo version as, in the spirit of this paper, it is applicable to arbitrary dimension whereas the efficient version only applies to surfaces embedded in $\mathbb{R}^3$.

\bibliographystyle{alpha}
\bibliography{main}

\newcommand{\etalchar}[1]{$^{#1}$}
\begin{thebibliography}{MMASC14}

\bibitem[BLM18]{Buet2018}
Blanche Buet, Gian Leonardi, and Simon Masnou.
\newblock Discretization and approximation of surfaces using varifolds.
\newblock {\em Geometric Flows}, 3:28--56, 03 2018.

\bibitem[CCSLT08]{chazal2008stability}
Frédéric Chazal, David Cohen-Steiner, André Lieutier, and Boris Thibert.
\newblock Stability of curvature measures, 2008.

\bibitem[CCSM11]{geo_inference_for_measures}
Frédéric Chazal, David Cohen-Steiner, and Quentin Mérigot.
\newblock Geometric inference for probability measures.
\newblock {\em Foundations of Computational Mathematics}, 11:733--751, 12 2011.

\bibitem[CLMT15]{Cuel_2015}
Louis Cuel, Jacques-Olivier Lachaud, Quentin M{\'{e} }rigot, and Boris Thibert.
\newblock Robust geometry estimation using the generalized voronoi covariance
  measure.
\newblock {\em {SIAM} Journal on Imaging Sciences}, 8(2):1293--1314, jan 2015.

\bibitem[CLS{\etalchar{+}}21]{Cao_2021}
Yueqi Cao, Didong Li, Huafei Sun, Amir~H. Assadi, and Shiqiang Zhang.
\newblock Efficient weingarten map and curvature estimation on manifolds.
\newblock {\em Machine Learning}, 110(6):1319--1344, may 2021.

\bibitem[CSM06]{cohen2006}
David Cohen-Steiner and J.~M. Morvan.
\newblock Second fundamental measure of geometric sets and local approximation
  of curvatures.
\newblock {\em Journal of Differential Geometry}, 74, 11 2006.

\bibitem[HDD{\etalchar{+}}92]{hoppe1992surface}
Hugues Hoppe, Tony DeRose, Tom Duchamp, John McDonald, and Werner Stuetzle.
\newblock Surface reconstruction from unorganized points.
\newblock In {\em Proceedings of the 19th annual conference on computer
  graphics and interactive techniques}, pages 71--78, 1992.

\bibitem[HKL20]{DBLP:journals/corr/abs-2001-07884}
Yuchen He, Sung~Ha Kang, and Hao Liu.
\newblock Curvature regularized surface reconstruction from point cloud.
\newblock {\em CoRR}, abs/2001.07884, 2020.

\bibitem[KBP{\etalchar{+}}04]{Koch2004C02}
Grady Koch, Bruce Barnes, Mulugeta Petros, Jeffrey Beyon, Farzin Amzajerdian,
  Jirong yu, Richard Davis, Syed Ismail, Stephanie Vay, Michael Kavaya, and
  Upendra Singh.
\newblock Coherent differential absorption lidar measurements of co2.
\newblock {\em Applied optics}, 43:5092--9, 10 2004.

\bibitem[LT19]{DBLP:journals/corr/abs-1910-13122}
Hazel Si~Min Lim and Araz Taeihagh.
\newblock Algorithmic decision-making in avs: Understanding ethical and
  technical concerns for smart cities.
\newblock {\em CoRR}, abs/1910.13122, 2019.

\bibitem[MMASC14]{monera2014taylor}
Maria~G Monera, A~Montesinos-Amilibia, and Esther Sanabria-Codesal.
\newblock The taylor expansion of the exponential map and geometric
  applications.
\newblock {\em Revista de la Real Academia de Ciencias Exactas, Fisicas y
  Naturales. Serie A. Matematicas}, 108:881--906, 2014.

\bibitem[MOG11]{merigotVORONOI2011}
Quentin Mérigot, Maks Ovsjanikov, and Leonidas~J. Guibas.
\newblock Voronoi-based curvature and feature estimation from point clouds.
\newblock {\em IEEE Transactions on Visualization and Computer Graphics},
  17(6):743--756, 2011.

\bibitem[MSR07]{MAGID2007139}
Evgeni Magid, Octavian Soldea, and Ehud Rivlin.
\newblock A comparison of gaussian and mean curvature estimation methods on
  triangular meshes of range image data.
\newblock {\em Computer Vision and Image Understanding}, 107(3):139--159, 2007.

\bibitem[Spi05]{spivak_2005}
Michael Spivak.
\newblock {\em A comprehensive introduction to differential geometry}.
\newblock Publish or Perish, Inc., 2005.

\end{thebibliography}

\end{document}